\documentclass[aps,prl,reprint,superscriptaddress,showpacs]{revtex4-1}
\usepackage{amsmath}
\usepackage{graphicx}
\usepackage{braket}
\usepackage{color}

\begin{document}

\title{Projected dipole moments of individual two-level defects extracted using circuit quantum electrodynamics}


\author{B. Sarabi}
\affiliation{Laboratory for Physical Sciences, College Park, MD 20740, USA}
\affiliation{Department of Physics, University of Maryland, College Park, MD 20742, USA}
\author{A. N. Ramanayaka}
\affiliation{Laboratory for Physical Sciences, College Park, MD 20740, USA}
\affiliation{Department of Physics, University of Maryland, College Park, MD 20742, USA}
\author{A. L. Burin}
\affiliation{Department of Chemistry, Tulane University, New Orleans, LA 70118, USA}
\author{F. C. Wellstood}
\affiliation{Department of Physics, University of Maryland, College Park, MD 20742, USA}
\affiliation{Joint Quantum Institute, University of Maryland, College Park, MD 20742, USA}
\author{K. D. Osborn}
\affiliation{Laboratory for Physical Sciences, College Park, MD 20740, USA}
\affiliation{Joint Quantum Institute, University of Maryland, College Park, MD 20742, USA}

	
\date{\today}
\begin{abstract}
Material-based two-level systems (TLSs), appearing as defects in low-temperature devices including superconducting qubits and photon detectors, are difficult to characterize. In this study we apply a uniform dc-electric field across a film to tune the energies of TLSs within. The film is embedded in a superconducting resonator such that it forms a circuit quantum electrodynamical (cQED) system. The energy of individual TLSs is observed as a function of the known tuning field.  By studying TLSs for which we can determine the tunneling energy, the {\em actual} $p_z$, dipole moments projected along the uniform field direction, are individually obtained. A distribution is created with 60 $p_z$.  We describe the distribution using a model with two dipole moment magnitudes, and a fit yields the corresponding values $p=p_1= 2.8\pm 0.2$ Debye and $p=p_2=8.3\pm0.4$ Debye. For a strong-coupled TLS the vacuum-Rabi splitting can be obtained with $p_z$ and tunneling energy. This allows a measurement of the circuit's zero-point electric field fluctuations, in a method that does not need the electric-field volume.
\end{abstract}


\maketitle
Dielectric two-level systems (TLSs) have attracted the attention of the quantum computing community ever since they were identified as a major source of decoherence in superconducting qubits \cite{PhysRevLett.95.210503}. Subsequent studies found that TLSs were also a performance-limiting source of noise in photon detectors used for astronomy \cite{GaoMKIDnoise, Mazin}. This motivation has led to quantum characterization of TLSs in the tunneling barrier of superconducting qubits \cite{PhysRevLett.105.177001, PhysRevLett.95.210503} and both noise and loss characterization in high quality superconducting resonator circuits. Thus TLSs are found as defects in various dielectric structures: deposited insulating films \cite{:/content/aip/journal/apl/92/11/10.1063/1.2898887,PhysRevLett.95.210503,PaikOsborn}, Josephson Junction (JJ) tunneling barriers \cite{PhysRevB.74.100502,PhysRevLett.95.210503,PhysRevB.78.144506}, imperfect interfaces between superconducting films with crystalline substrates \cite{:/content/aip/journal/apl/97/23/10.1063/1.3517252}, and the native oxides on materials \cite{:/content/aip/journal/jap/114/5/10.1063/1.4817512}. Recent modeling has predicted possible structures and values for the TLS dipole moment \cite{PhysRevLett.111.065901,gordon2014hydrogen,PhysRevLett.110.077002}. While these TLSs are generally known to be charged atomic configurations that spatially tunnel, their microscopic structure and elemental composition are generally unknown.

In a qubit made from an anharmonic oscillator the interaction energy is observed as a spectroscopic splitting in the qubit state. However, this quantity is not only dependent on circuit element parameters, but also on the thickness of the tunneling barrier \cite{PhysRevLett.95.210503}. The barrier has a thickness variation of 1-2 nm \cite{BarrierVary}, such that this lack of accuracy is also present in the electric field amplitude and the measured TLS dipole moment. In general, individual measurements of TLSs within JJs find the transition dipole moment $p_{tr}=p_z\Delta_{0}/\mathcal{E}$, where $p_z\equiv|p_z|$ is the absolute value of the dipole moment projected in the field direction, and the TLS tunneling energy over the total energy $\Delta_{0}/\mathcal{E}$ is generally unknown. As a result only a lower bound of $p_z$ is determined from those measurements ($\Delta_{0}/\mathcal{E}\leq1$).  However, in studies of many individual TLSs in the barrier of a JJ, the distribution of observations is consistent with a single TLS moment magnitude $p$ \cite{PhysRevLett.95.210503, Stoutimore}, assuming a standard distribution of $\Delta_{0}$ \cite{PhillipsJLTP,doi:10.1080/14786437208229210}. Here only the average projected moment $\braket{p_z}\approx p/\sqrt{3}$ is determined from the model fit, and has an accuracy limitation caused by the aforementioned variable barrier thickness.

TLSs have been recently studied with an applied field which tilts the TLS potential energies. A strain tuning study with individual TLS observations \cite{Grabovskij12102012} verified agreement with the standard model which employs double-well potentials \cite{PhillipsJLTP,doi:10.1080/14786437208229210}. Further work revealed TLS-TLS interactions \cite{lisenfeld}.  Finally, a swept electric-field study measured nonequilibrium microwave loss which was explained by Landau-Zener dynamics and a single representative dipole moment (magnitude) \cite{khalil2013landau,PhysRevLett.110.157002}.

\begin{figure}
\includegraphics[width=\linewidth]{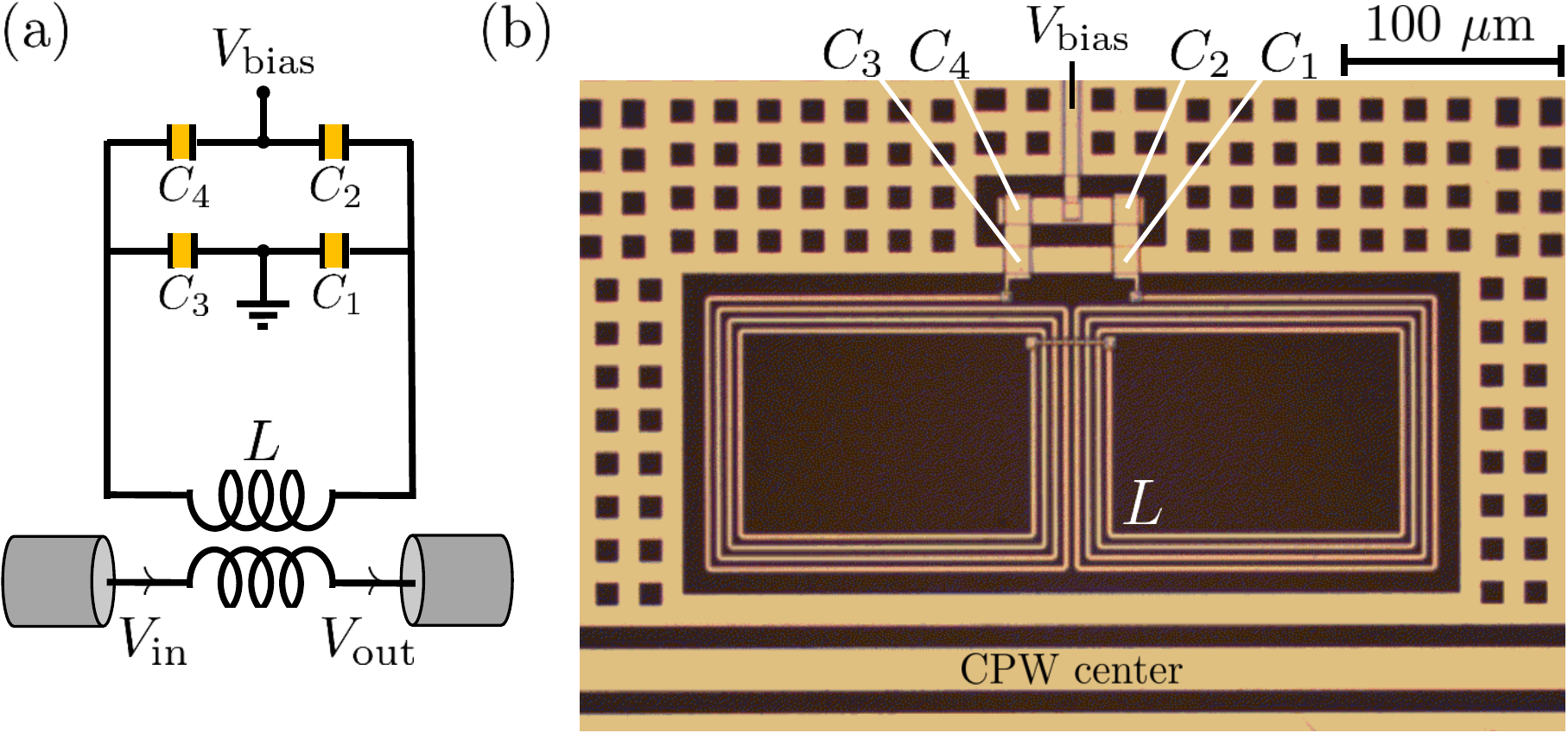}
\caption{(a) Schematic of LC resonator coupled to TLSs within the capacitors C1-C4 of the electric field controlled device. The voltage $V_{bias}$ induces a uniform dc-electric field across the capacitors. (b) Optical image of the fabricated tunable device. Aluminum appears light and the sapphire substrate appears black. }
\label{opticalimage}
\end{figure}

\begin{figure*}
\includegraphics[width=\textwidth]{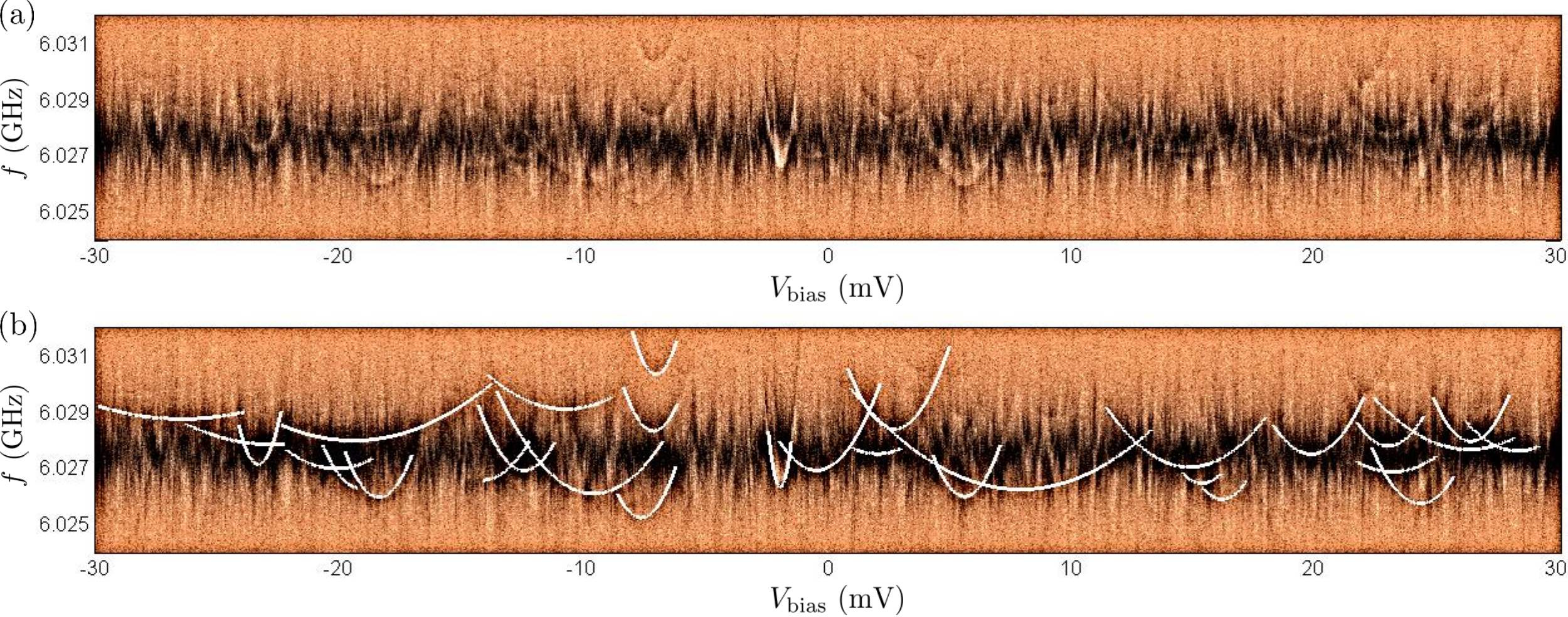}
\caption{(a) False-color plot of transmission $|S_{21}|=|V_{out}/V_{in}|$ vs. frequency $f$ and bias voltage $V_{\mathrm{bias}}$. Data is taken at $T=25$ mK and $\bar{n}_{\mathrm{max}}\simeq0.4$. Light copper and black correspond to $|S_{21}|=0.56$ and $|S_{21}|=0.40$, respectively. TLSs are observed with minima in energy (the tunneling energy $\Delta_{0}$) nearly degenerate with the circuit photon energy $\hbar\omega_{c}$. (b) Fits with energy model (Eq. 1) to data in panel (a) are shown which find the projected moment $p_z$ and tunneling energy $\Delta_0$ of 30 TLSs.}
\label{spectroscopy}
\end{figure*}

In our device, TLSs are coupled to a LC resonator and tuned with a {\em uniform and known dc electric field $\mathbf{E}_{bias}$}. This tuning distinguishes it from an earlier study of the vacuum Rabi splitting and Glauber state made with a TLS \cite{sarabi2014cavity}. TLSs are chosen for analysis only near an observed energy minima, which reveals the tunneling energies $\Delta_{0}$. In our experiment, $p_{z}\equiv p|\cos\theta|$ is measured for each individual TLS and enabled by that of $\Delta_0$. This leaves the relationship to the dipole moment magnitude $p$ uncertain only by the angle $\theta$ between the dipole and the known dc-field direction. 60 $p_z$ are individually extracted. The distribution created from these values is fit to a model with two dipole moment magnitudes, $p=p_1$ and $p=p_2$, and thus is related to earlier studies using two loss tangent sources \cite{Sandberg, ALDOxides}. While $p_{tr}$ can be determined in general from dc-tuned measurements, some TLSs also have strong-coupling such that the vacuum Rabi splitting is observed with the corresponding coupling frequency $g=p_{tr}E_{RMS}/\hbar$. By combining a measurement of $p_z$, $\Delta_{0}$ and $g$ we get a complete measurement of the uniform zero-point electric field fluctuations $E_{RMS}$ of the device. As will be discussed below, our film contains TLSs related to those from JJ barriers, material interfaces, contamination, and dielectrics in general, such that the accurate moments can generally be extracted by this type of device.

In this experiment we use a superconducting resonator with bridge arrangement of four parallel-plate capacitors shown in Fig. \ref{opticalimage}. The capacitors allow a uniform dc electric bias $\mathbf{E}_{\mathrm{bias}}$ to the TLSs within, creating the transition energy
\begin{equation}\label{TLStune}
\mathcal{E}=\sqrt{\Delta_{0}^{2}+(\Delta+2p_zE_{\mathrm{bias}})^{2}}.
\end{equation} Here, $\Delta$ is the energy difference between the potential well minima. The tunneling dipole can alternatively be described with a distance between the wells $2p/q$ for a tunneling charge $q$. The TLS distribution from the standard model is $dn^2\propto d\Delta_{0}d\Delta/\Delta_{0}$ \cite{PhillipsJLTP,doi:10.1080/14786437208229210,anderson1978lectures, PhysRevB.80.172506,:/content/aip/journal/apl/97/25/10.1063/1.3529457}. Through symmetry the bias lead at voltage $V_{\mathrm{bias}}$ is isolated from the ac resonance. The capacitors are formed with two aluminum films and a $d_{0}=125$ nm thick layer of silicon nitride (SiN$_{\mathrm{x}}$) in between. Since the bridge capacitance contains two capacitors in series, the TLSs within experience the bias field $E_{\mathrm{bias}}=V_{\mathrm{bias}}/(2 d_0)$.  The capacitance and a quadrupole spiral inductor produce a resonance at 6.0 GHz. The SiN$_{\mathrm{x}}$ dielectric volume within the capacitors $V=78\:\mu\mathrm{m}^{3}$ is micron-scaled (micro-$V$) to allow strong enough coupling for individual TLS observations.

The dielectric used in this study has a loss tangent of $\tan \delta_0=7.8\times10^{-4}$ as obtained from the $p_{z}$ data below \cite{Supplemental}.  The silicon nitride is deposited with PECVD\cite{PaikOsborn}. Individual TLS have also been observed using films with smaller $\tan \delta_0$ \cite{sarabi2014cavity}, but the film for this study was selected to allow many TLSs to be observed in a small range of bias voltage $V_{\mathrm{bias}}$. The microwave input power was generally set such that the photon occupancy on resonance is $\bar{n}_{\mathrm{max}}\simeq0.4$. This photon occupancy allows sufficient signal to noise, and in a similar device is found to resolve nearly degenerate resonator-TLS energy levels in the presence of other weaker coupled TLSs \cite{Bahmanthesis}.

The micro-$V$ device was measured at a temperature $T\leq25$ mK in three data sets from separate cool-downs. In each data set the bias voltage $V_{\mathrm{bias}}$ was varied by 60 mV across the dielectric corresponding to an electric field range of 240 kV/m within it. The bias line was low-pass filtered at room temperature and on the 3 Kelvin stage to reduce noise induced to the TLS energies. In addition, a 3 dB microwave attenuator was installed on the 0.7 K plate to thermalize the center conductor of the corresponding coaxial cable. Finally, a Cu-powder low-pass filter was used at the base temperature stage of the refrigerator to absorb Cooper-pair breaking photons.

\begin{figure}
\includegraphics[trim = 42mm 90mm 30mm 95mm, clip, width=10cm]{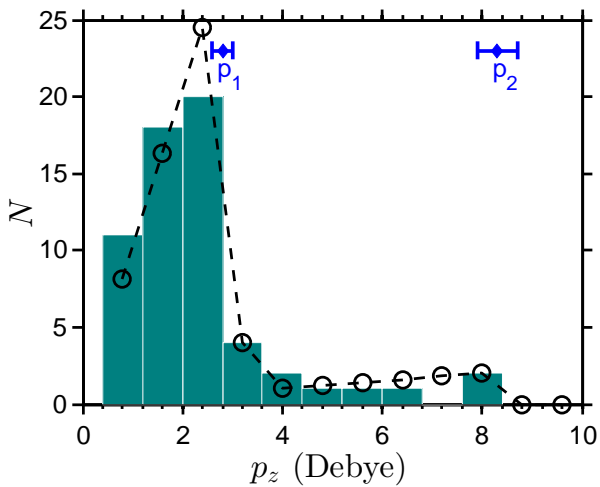}
\caption{Distribution of $p_{z}$ as determined from fits similar to Fig. 2 (b). An initial fit to the model with two dipole moment magnitudes $p$ is shown in circle markers and dashed lines. The two-sawtooth pattern follows from the technique and a standard model of random TLS orientations. The most likely ranges for moments magnitudes, $p_1= 2.8\pm 0.2$ Debye and $p_2=8.3\pm0.4$ Debye, are shown by the markers with horizontal error bars.}
\label{histogram}
\end{figure}

Figure \ref{spectroscopy}(a) shows the transmission spectroscopy results from one of the data sets. The wide dark region generally corresponds to the broad circuit resonance $\omega_{c}/2\pi$, and a large number of fine features are caused by near-resonance interactions with individual TLSs. Using a first analysis which neglects vacuum splitting effects, we interpret the fine features as the TLS energies. In agreement with the standard model for energy Eq. \ref{TLStune}, some TLSs exhibit a minimum in their energy (at $\hbar\omega_{c} \simeq \Delta_{0}$) versus $E_{bias}$. The energy is also symmetric about the minimum in the same quantity, as expected from the referenced equation. Since $E_{bias}$ is directly applied, a fit to the data, as shown in Fig. \ref{spectroscopy}(b), produces a fairly direct determination of the individual projected dipole moments $p_z$. Each $p_z$ is extracted with a precision estimated at approximately 1 percent (see below). $\Delta=-p_z V_{bias}/d_0$ when the energy is at a minimum ($\mathcal{E}=\Delta_{0}$). At this bias, degnerate wells (DW) are created in the TLS double well potential (as well as the even-amplitude superposition of the single well states).

Fig. 3 shows the distribution of 60 $p_z$ extracted using data from three separate cool downs. The resulting distribution of $p_z$ can be interpreted to have two peaks, and suggests that we compare it to the standard amorphous model of TLSs containing two dipole-moment magnitudes, $p_1$ and $p_2$. In accordance with the random-orientation defect model, we write the TLS number density $d^3n=dp_z d\Delta d\Delta_{0} (P_1 \Theta(p_1-p_z)/(p_1 \Delta_{0})+P_2 \Theta(p_2-p_z)/(p_2 \Delta_{0}))$ in terms of the projected dipole moment $p_z$, where $\Theta$ is the Heaviside step function and $P_i$ is the material constant for $p_i$, and $i = 1, 2$ \cite{Supplemental}. The range of dc-bias creates an observable set of DWs with a distribution
\begin{equation}\label{DWS}
\frac{dn_{DW}}{dp_z}= \frac{\kappa}{\omega_c} \frac{p_z\delta V_{\mathrm{bias}}}{d_0} (\frac{P_1 \Theta(p_1-p_z)}{p_1}+\frac{P_2 \Theta(p_2-p_z)}{p_2}).
\end{equation}
Here the experimental ranges are $\delta\Delta_0=\hbar\kappa$ and $\delta\Delta=p_z\delta V_{bias}/d_0$, where $\kappa$ is the coupling rate between the resonator and transmission line. We see from theory that for each $p_{i}$ there should be a number of degenerate-well TLS in a bin $N$, where $N \propto p_z$. This feature is a consequence of how larger $p_z$ will be swept further in energy for a fixed domain in $\delta V_{\mathrm{bias}}$.

An initial fit of Eq. \ref{DWS} to the data yields $p_1=2.9$ Debye and $p_2=8.4$ Debye, as shown in the figure, where circle markers and the dashed line indicate bin heights. The fit to the data is consistent with the model if the bin to the right of 2.8 Debye has some counts from the $p_1$ distribution while the bin to the left of 2.8 Debye has lower counts than the fit value due to Poisson statistical fluctuations. Using $P_1$ and $P_2$ from the initial fit, Monte Carlo analysis was performed using different input values for $p_1$ and $p_2$. Simulation datasets drew the statistically correct number of $p_z$ in simulation distributions. Fits with optimization for Poisson statistics \cite{Poisson} were then used. These fits created simulation-extracted $p_1$ and $p_2$ which generally differ from input values due to Poisson sampling error. The difference was used to find the probable range of $p_1$ and $p_2$ for a two-moment material. Hundreds of simulated datasets were used to find the range of moments consistent with the observed data. The analysis gives $p_1= 2.8\pm 0.2$ Debye and $p_2=8.3\pm0.4$ Debye for the moments of the model, where the range indicates a single standard deviation. Thus we find our data is consistent with a model of two dipole moments, and that with the use of this model the standard deviations of dipole moments are $\le$7$\%$.

To the knowledge of the authors, this experiment marks the first study of a single sample from which two TLS moment magnitudes were studied. In applications with this material we expect that $p_1$ is the more important moment for material characterization since TLSs in this part of the distribution are much more numerous. In an alumina barrier of a JJ qubit a TLS dipole moment of 3 Debye was found if one assumes the barrier thickness of 2 nm \cite{PhysRevLett.95.210503} (TLSs detected in tunneling barriers are dependent on an unknown barrier thickness). In $\delta_0$-averaged measurements of bulk SiO$_{2}$ with OH$^{-}$ correlated TLSs the moment magnitude extracted is 3.3 Debye \cite{PhysRevB.38.9952}. $p_1$ of our nitride film is similar and, from SIMS analysis, our films are also known to contain some O and H. Although our value of $p_2$ is large, it is within the range of values reported in other measurements of silicon nitride \cite{khalil2013landau}. We believe that the DW extraction technique creates the potential for quantitative analysis on samples, where different dipole distribution models could be tested with statistical analysis.

Furthermore, since our method uses an insulating (current blocking) layer for TLS, it can be extended for use to characterize various materials which are important to quantum-sensitive devices. Any insulating film can be tested using this technique, and this includes films of alumina, the material commonly used in JJ tunneling barriers. In addition, the technique can accommodate multiple layers of TLS-laden films, such that it could test surface oxides, states at a superconductor/insulator boundary, process contamination at the same boundary, the thickness dependence of TLSs in films, etc. Finally, our technique could test the dependence of $p$ on an independent variable, such as hydrogen concentration at a host material or interface.

\begin{figure}
\includegraphics[width=\linewidth]{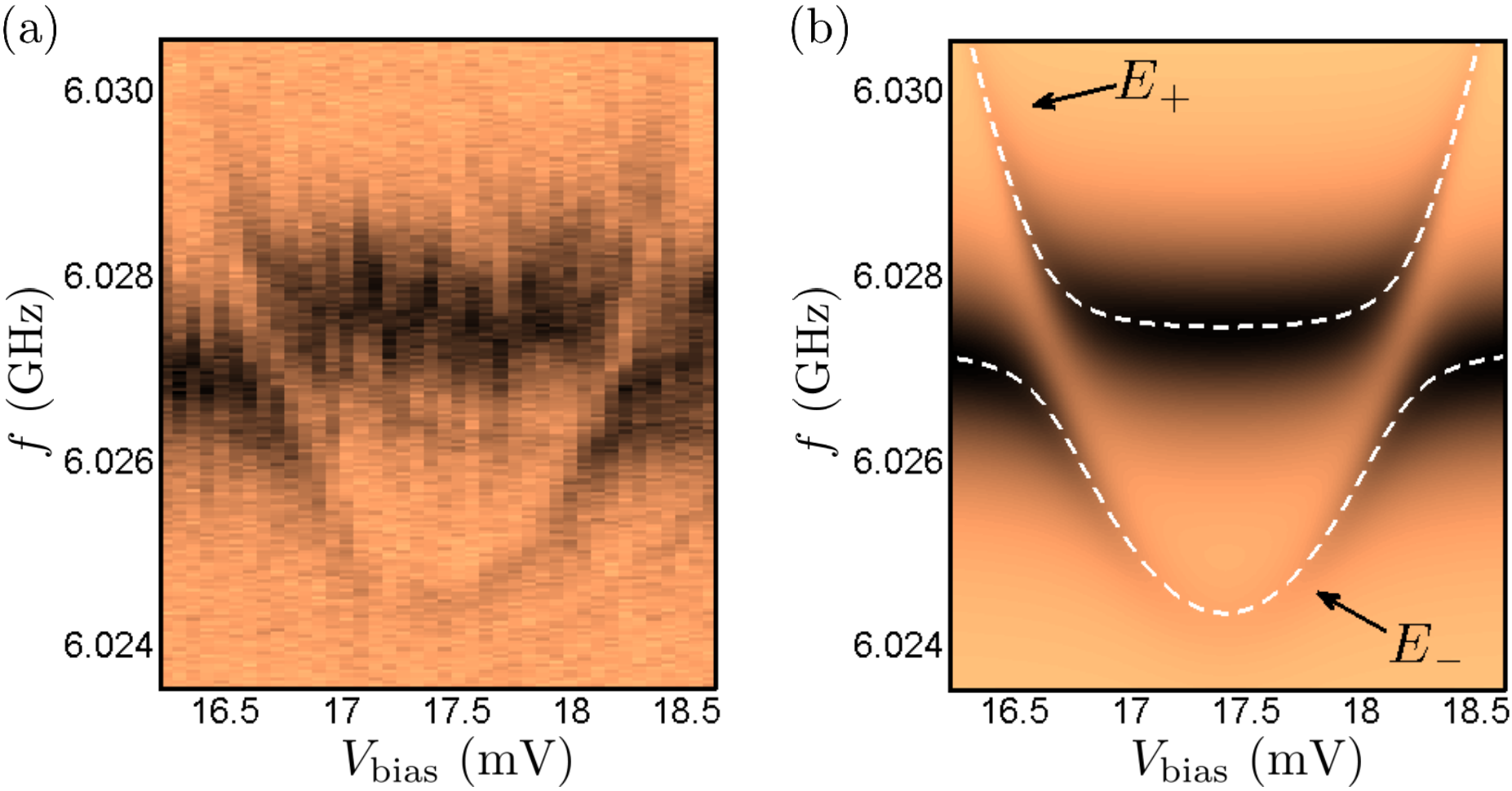}
\caption{(a) False-color plot of measured $|S_{21}|$ vs. frequency $f$ and bias voltage $V_{\mathrm{bias}}$ for one of the most strongly coupled TLSs. The resonator-TLS coupling induces avoided crossings dependent on the coupling $g$. Data is taken at $T=24$ mK and a maximum photon occupancy of $\bar{n}_{\mathrm{max}}\simeq0.4$. Light copper and black correspond to $|S_{21}|=0.60$ and $|S_{21}|=0.43$, respectively. (b) Optimum fit to panel (a) data using a cQED-based model (see main text). White dashed curves show the extracted low-power eigenenergies $E_{+}$ and $E_{-}$ of the system. $g$ is extracted along with $p_z$ and $\Delta_0$, without the electric volume.}
\label{fitting}
\end{figure}

We next analyze one of the most strongly coupled TLS to the resonator, where Fig. \ref{fitting}(a) shows a close-up of the DW spectra.  Here we clearly see the minima energy of the TLS with the vacuum Rabi splitting. The Jaynes-Cummings eigenenergies for this system are
\begin{equation}\label{eq:transitionenergiesfitting}
E_{\pm}=\frac{1}{2}\hbar(\omega_{c}+\omega_{\mathrm{TLS}})\pm\hbar\sqrt{g^{2}+(\delta/2)^{2}},
\end{equation}
where $\omega_{\mathrm{TLS}}=\mathcal{E}/\hbar$ (Eq. 1), and $\delta$ represents the TLS-resonator detuning. Figure \ref{fitting}(b) shows the least squares Monte Carlo (LSM) fit to the spectra using the theory of their lineshapes Ref. \cite{sarabi2014cavity, Bahmanthesis}. The eigenenergies from the fit determined $p_{z}=6.0$ Debye  and coupling $g/2\pi=753$ kHz are shown as white dashed curves. The fit also yields the coherence time T$_{2}=2/\gamma_{\mathrm{TLS}}=313$ ns, related to the spectral width. Here, $\gamma_{\mathrm{TLS}}$ is the TLS relaxation rate which limits coherence because the temperature is low and strong coupling to the environment, through the cavity, provides enhanced relaxation. For this TLS the eigenenergy exhibits a visible splitting ($g$) such that this fit yields a more accurate $p_z$ than extracted from the previous method, e.g., the method of Fig. 2(b). However, the difference in $p_z$ caused by even the strongest cQED coupling is small; $p_z$ is only different by 1$\%$ between the two methods. This shows an example of how $p_z$ extracted earlier (for Fig. \ref{histogram}) using a weak-coupling assumption are only weakly perturbed. Since $g/\kappa=1.7$ is larger than 1 we are in the strong coupling regime. However, the TLS relaxation is large in this case, $g/\gamma_{\mathrm{TLS}}=0.74$, showing that the strong coupling limit does not rely on good TLS coherence \cite{Kimble}.

This analysis also gives a measurement of the RMS fluctuation field $E_{\mathrm{RMS}}$ without using the geometrically estimated electric-field volume. From the optimum fit parameters we calculate $E_{\mathrm{RMS}}=\hbar g\mathcal{E}/p_{z}\Delta_{0}=25$ V/m (where $\Delta_{0}/\mathcal{E}\simeq1$). On the other hand, the design parameters allow us to calculate $E_{\mathrm{RMS}}=\hbar\sqrt{\omega_{c}/(2\epsilon_{r}\epsilon_{0}\hbar V)}=21$ V/m (using $\epsilon_{r}=6.5$ \cite{PaikOsborn} and the bridge capacitors' dielectric volume $V$), which is 16\% smaller than the fit determined $E_{\mathrm{RMS}}$. Since $V$ was not underestimated and $p_{z}$ is very accurate, the difference is believed to be caused by the uncertainty in the measured $g$.

In conclusion, a resonator was studied which accurately applies a uniform dc-electric field to TLSs in a film. The field allows TLS energies to be analyzed as a function of a known field and the direct extraction of 60 projected moments $p_z$. An analysis reveals a good fit to a model with two representative dipole moment magnitudes for a silicon nitride film, $p_1= 2.8\pm 0.2 $ and $p_2=8.3\pm0.4$ Debye. The $p_1$ of our nitride film was comparable to that of OH$^{-}$ found in bulk silicon oxide. With additional data this method should allow one to search for additional dipole magnitudes (TLS types). Since this study uses a deposited insulating film containing TLSs, it could be extended to analyze other materials and structures (surface oxides, dielectric-superconductor interface states, fabrication contamination, interdigital capacitors, etc.). This should allow the classification of TLSs by dipole moment and the optimization of quantum computing or photon detector devices. The cQED measurement also allowed a complete measurement of the zero-point fluctuation electric field. This demonstrates how strong-coupling cQED can be achieved with TLSs in a practical energy-tunable circuit such that many measurements which are available for superconducting qubits should now be available for TLSs.

The authors would like to thank Christopher Lobb, Ray Simmonds and Yaniv Rosen for many useful discussions.


\begin{thebibliography}{34}%
\makeatletter
\providecommand \@ifxundefined [1]{%
 \@ifx{#1\undefined}
}%
\providecommand \@ifnum [1]{%
 \ifnum #1\expandafter \@firstoftwo
 \else \expandafter \@secondoftwo
 \fi
}%
\providecommand \@ifx [1]{%
 \ifx #1\expandafter \@firstoftwo
 \else \expandafter \@secondoftwo
 \fi
}%
\providecommand \natexlab [1]{#1}%
\providecommand \enquote  [1]{``#1''}%
\providecommand \bibnamefont  [1]{#1}%
\providecommand \bibfnamefont [1]{#1}%
\providecommand \citenamefont [1]{#1}%
\providecommand \href@noop [0]{\@secondoftwo}%
\providecommand \href [0]{\begingroup \@sanitize@url \@href}%
\providecommand \@href[1]{\@@startlink{#1}\@@href}%
\providecommand \@@href[1]{\endgroup#1\@@endlink}%
\providecommand \@sanitize@url [0]{\catcode `\\12\catcode `\$12\catcode
  `\&12\catcode `\#12\catcode `\^12\catcode `\_12\catcode `\%12\relax}%
\providecommand \@@startlink[1]{}%
\providecommand \@@endlink[0]{}%
\providecommand \url  [0]{\begingroup\@sanitize@url \@url }%
\providecommand \@url [1]{\endgroup\@href {#1}{\urlprefix }}%
\providecommand \urlprefix  [0]{URL }%
\providecommand \Eprint [0]{\href }%
\providecommand \doibase [0]{http://dx.doi.org/}%
\providecommand \selectlanguage [0]{\@gobble}%
\providecommand \bibinfo  [0]{\@secondoftwo}%
\providecommand \bibfield  [0]{\@secondoftwo}%
\providecommand \translation [1]{[#1]}%
\providecommand \BibitemOpen [0]{}%
\providecommand \bibitemStop [0]{}%
\providecommand \bibitemNoStop [0]{.\EOS\space}%
\providecommand \EOS [0]{\spacefactor3000\relax}%
\providecommand \BibitemShut  [1]{\csname bibitem#1\endcsname}%
\let\auto@bib@innerbib\@empty
\bibitem [{\citenamefont {Martinis}\ \emph {et~al.}(2005)\citenamefont
  {Martinis}, \citenamefont {Cooper}, \citenamefont {McDermott}, \citenamefont
  {Steffen}, \citenamefont {Ansmann}, \citenamefont {Osborn}, \citenamefont
  {Cicak}, \citenamefont {Oh}, \citenamefont {Pappas}, \citenamefont
  {Simmonds},\ and\ \citenamefont {Yu}}]{PhysRevLett.95.210503}%
  \BibitemOpen
  \bibfield  {author} {\bibinfo {author} {\bibfnamefont {J.~M.}\ \bibnamefont
  {Martinis}}, \bibinfo {author} {\bibfnamefont {K.~B.}\ \bibnamefont
  {Cooper}}, \bibinfo {author} {\bibfnamefont {R.}~\bibnamefont {McDermott}},
  \bibinfo {author} {\bibfnamefont {M.}~\bibnamefont {Steffen}}, \bibinfo
  {author} {\bibfnamefont {M.}~\bibnamefont {Ansmann}}, \bibinfo {author}
  {\bibfnamefont {K.~D.}\ \bibnamefont {Osborn}}, \bibinfo {author}
  {\bibfnamefont {K.}~\bibnamefont {Cicak}}, \bibinfo {author} {\bibfnamefont
  {S.}~\bibnamefont {Oh}}, \bibinfo {author} {\bibfnamefont {D.~P.}\
  \bibnamefont {Pappas}}, \bibinfo {author} {\bibfnamefont {R.~W.}\
  \bibnamefont {Simmonds}}, \ and\ \bibinfo {author} {\bibfnamefont {C.~C.}\
  \bibnamefont {Yu}},\ }\href {\doibase 10.1103/PhysRevLett.95.210503}
  {\bibfield  {journal} {\bibinfo  {journal} {Phys. Rev. Lett.}\ }\textbf
  {\bibinfo {volume} {95}},\ \bibinfo {pages} {210503} (\bibinfo {year}
  {2005})}\BibitemShut {NoStop}%
\bibitem [{\citenamefont {Gao}\ \emph {et~al.}(2007)\citenamefont {Gao},
  \citenamefont {Zmuidzinas}, \citenamefont {Mazin}, \citenamefont {LeDuc},\
  and\ \citenamefont {Day}}]{GaoMKIDnoise}%
  \BibitemOpen
  \bibfield  {author} {\bibinfo {author} {\bibfnamefont {J.}~\bibnamefont
  {Gao}}, \bibinfo {author} {\bibfnamefont {J.}~\bibnamefont {Zmuidzinas}},
  \bibinfo {author} {\bibfnamefont {B.~A.}\ \bibnamefont {Mazin}}, \bibinfo
  {author} {\bibfnamefont {H.~G.}\ \bibnamefont {LeDuc}}, \ and\ \bibinfo
  {author} {\bibfnamefont {P.~K.}\ \bibnamefont {Day}},\ }\href {\doibase
  http://dx.doi.org/10.1063/1.2711770} {\bibfield  {journal} {\bibinfo
  {journal} {Applied Physics Letters}\ }\textbf {\bibinfo {volume} {90}},\
  \bibinfo {eid} {102507} (\bibinfo {year} {2007})}\BibitemShut {NoStop}%
\bibitem [{\citenamefont {Mazin}(2009)}]{Mazin}%
  \BibitemOpen
  \bibfield  {author} {\bibinfo {author} {\bibfnamefont {B.~A.}\ \bibnamefont
  {Mazin}},\ }\href@noop {} {\bibfield  {journal} {\bibinfo  {journal} {AIP
  Conference Proceedings}\ }\textbf {\bibinfo {volume} {1185}} (\bibinfo {year}
  {2009})}\BibitemShut {NoStop}%
\bibitem [{\citenamefont {Shalibo}\ \emph {et~al.}(2010)\citenamefont
  {Shalibo}, \citenamefont {Rofe}, \citenamefont {Shwa}, \citenamefont
  {Zeides}, \citenamefont {Neeley}, \citenamefont {Martinis},\ and\
  \citenamefont {Katz}}]{PhysRevLett.105.177001}%
  \BibitemOpen
  \bibfield  {author} {\bibinfo {author} {\bibfnamefont {Y.}~\bibnamefont
  {Shalibo}}, \bibinfo {author} {\bibfnamefont {Y.}~\bibnamefont {Rofe}},
  \bibinfo {author} {\bibfnamefont {D.}~\bibnamefont {Shwa}}, \bibinfo {author}
  {\bibfnamefont {F.}~\bibnamefont {Zeides}}, \bibinfo {author} {\bibfnamefont
  {M.}~\bibnamefont {Neeley}}, \bibinfo {author} {\bibfnamefont {J.~M.}\
  \bibnamefont {Martinis}}, \ and\ \bibinfo {author} {\bibfnamefont
  {N.}~\bibnamefont {Katz}},\ }\href {\doibase 10.1103/PhysRevLett.105.177001}
  {\bibfield  {journal} {\bibinfo  {journal} {Phys. Rev. Lett.}\ }\textbf
  {\bibinfo {volume} {105}},\ \bibinfo {pages} {177001} (\bibinfo {year}
  {2010})}\BibitemShut {NoStop}%
\bibitem [{\citenamefont {O`Connell}\ \emph {et~al.}(2008)\citenamefont
  {O`Connell}, \citenamefont {Ansmann}, \citenamefont {Bialczak}, \citenamefont
  {Hofheinz}, \citenamefont {Katz}, \citenamefont {Lucero}, \citenamefont
  {McKenney}, \citenamefont {Neeley}, \citenamefont {Wang}, \citenamefont
  {Weig}, \citenamefont {Cleland},\ and\ \citenamefont
  {Martinis}}]{:/content/aip/journal/apl/92/11/10.1063/1.2898887}%
  \BibitemOpen
  \bibfield  {author} {\bibinfo {author} {\bibfnamefont {A.~D.}\ \bibnamefont
  {O`Connell}}, \bibinfo {author} {\bibfnamefont {M.}~\bibnamefont {Ansmann}},
  \bibinfo {author} {\bibfnamefont {R.~C.}\ \bibnamefont {Bialczak}}, \bibinfo
  {author} {\bibfnamefont {M.}~\bibnamefont {Hofheinz}}, \bibinfo {author}
  {\bibfnamefont {N.}~\bibnamefont {Katz}}, \bibinfo {author} {\bibfnamefont
  {E.}~\bibnamefont {Lucero}}, \bibinfo {author} {\bibfnamefont
  {C.}~\bibnamefont {McKenney}}, \bibinfo {author} {\bibfnamefont
  {M.}~\bibnamefont {Neeley}}, \bibinfo {author} {\bibfnamefont
  {H.}~\bibnamefont {Wang}}, \bibinfo {author} {\bibfnamefont {E.~M.}\
  \bibnamefont {Weig}}, \bibinfo {author} {\bibfnamefont {A.~N.}\ \bibnamefont
  {Cleland}}, \ and\ \bibinfo {author} {\bibfnamefont {J.~M.}\ \bibnamefont
  {Martinis}},\ }\href {\doibase http://dx.doi.org/10.1063/1.2898887}
  {\bibfield  {journal} {\bibinfo  {journal} {Appl. Phys. Lett.}\ }\textbf
  {\bibinfo {volume} {92}},\ \bibinfo {eid} {112903} (\bibinfo {year}
  {2008})}\BibitemShut {NoStop}%
\bibitem [{\citenamefont {Paik}\ and\ \citenamefont
  {Osborn}(2010)}]{PaikOsborn}%
  \BibitemOpen
  \bibfield  {author} {\bibinfo {author} {\bibfnamefont {H.}~\bibnamefont
  {Paik}}\ and\ \bibinfo {author} {\bibfnamefont {K.~D.}\ \bibnamefont
  {Osborn}},\ }\href {\doibase http://dx.doi.org/10.1063/1.3309703} {\bibfield
  {journal} {\bibinfo  {journal} {Applied Physics Letters}\ }\textbf {\bibinfo
  {volume} {96}},\ \bibinfo {eid} {072505} (\bibinfo {year}
  {2010})}\BibitemShut {NoStop}%
\bibitem [{\citenamefont {Oh}\ \emph {et~al.}(2006)\citenamefont {Oh},
  \citenamefont {Cicak}, \citenamefont {Kline}, \citenamefont {Sillanp\"a\"a},
  \citenamefont {Osborn}, \citenamefont {Whittaker}, \citenamefont {Simmonds},\
  and\ \citenamefont {Pappas}}]{PhysRevB.74.100502}%
  \BibitemOpen
  \bibfield  {author} {\bibinfo {author} {\bibfnamefont {S.}~\bibnamefont
  {Oh}}, \bibinfo {author} {\bibfnamefont {K.}~\bibnamefont {Cicak}}, \bibinfo
  {author} {\bibfnamefont {J.~S.}\ \bibnamefont {Kline}}, \bibinfo {author}
  {\bibfnamefont {M.~A.}\ \bibnamefont {Sillanp\"a\"a}}, \bibinfo {author}
  {\bibfnamefont {K.~D.}\ \bibnamefont {Osborn}}, \bibinfo {author}
  {\bibfnamefont {J.~D.}\ \bibnamefont {Whittaker}}, \bibinfo {author}
  {\bibfnamefont {R.~W.}\ \bibnamefont {Simmonds}}, \ and\ \bibinfo {author}
  {\bibfnamefont {D.~P.}\ \bibnamefont {Pappas}},\ }\href {\doibase
  10.1103/PhysRevB.74.100502} {\bibfield  {journal} {\bibinfo  {journal} {Phys.
  Rev. B}\ }\textbf {\bibinfo {volume} {74}},\ \bibinfo {pages} {100502}
  (\bibinfo {year} {2006})}\BibitemShut {NoStop}%
\bibitem [{\citenamefont {Kim}\ \emph {et~al.}(2008)\citenamefont {Kim},
  \citenamefont {Zaretskey}, \citenamefont {Yoon}, \citenamefont
  {Schneiderman}, \citenamefont {Shaw}, \citenamefont {Echternach},
  \citenamefont {Wellstood},\ and\ \citenamefont
  {Palmer}}]{PhysRevB.78.144506}%
  \BibitemOpen
  \bibfield  {author} {\bibinfo {author} {\bibfnamefont {Z.}~\bibnamefont
  {Kim}}, \bibinfo {author} {\bibfnamefont {V.}~\bibnamefont {Zaretskey}},
  \bibinfo {author} {\bibfnamefont {Y.}~\bibnamefont {Yoon}}, \bibinfo {author}
  {\bibfnamefont {J.~F.}\ \bibnamefont {Schneiderman}}, \bibinfo {author}
  {\bibfnamefont {M.~D.}\ \bibnamefont {Shaw}}, \bibinfo {author}
  {\bibfnamefont {P.~M.}\ \bibnamefont {Echternach}}, \bibinfo {author}
  {\bibfnamefont {F.~C.}\ \bibnamefont {Wellstood}}, \ and\ \bibinfo {author}
  {\bibfnamefont {B.~S.}\ \bibnamefont {Palmer}},\ }\href {\doibase
  10.1103/PhysRevB.78.144506} {\bibfield  {journal} {\bibinfo  {journal} {Phys.
  Rev. B}\ }\textbf {\bibinfo {volume} {78}},\ \bibinfo {pages} {144506}
  (\bibinfo {year} {2008})}\BibitemShut {NoStop}%
\bibitem [{\citenamefont {Vissers}\ \emph {et~al.}(2010)\citenamefont
  {Vissers}, \citenamefont {Gao}, \citenamefont {Wisbey}, \citenamefont {Hite},
  \citenamefont {Tsuei}, \citenamefont {Corcoles}, \citenamefont {Steffen},\
  and\ \citenamefont
  {Pappas}}]{:/content/aip/journal/apl/97/23/10.1063/1.3517252}%
  \BibitemOpen
  \bibfield  {author} {\bibinfo {author} {\bibfnamefont {M.~R.}\ \bibnamefont
  {Vissers}}, \bibinfo {author} {\bibfnamefont {J.}~\bibnamefont {Gao}},
  \bibinfo {author} {\bibfnamefont {D.~S.}\ \bibnamefont {Wisbey}}, \bibinfo
  {author} {\bibfnamefont {D.~A.}\ \bibnamefont {Hite}}, \bibinfo {author}
  {\bibfnamefont {C.~C.}\ \bibnamefont {Tsuei}}, \bibinfo {author}
  {\bibfnamefont {A.~D.}\ \bibnamefont {Corcoles}}, \bibinfo {author}
  {\bibfnamefont {M.}~\bibnamefont {Steffen}}, \ and\ \bibinfo {author}
  {\bibfnamefont {D.~P.}\ \bibnamefont {Pappas}},\ }\href {\doibase
  http://dx.doi.org/10.1063/1.3517252} {\bibfield  {journal} {\bibinfo
  {journal} {Applied Physics Letters}\ }\textbf {\bibinfo {volume} {97}},\
  \bibinfo {eid} {232509} (\bibinfo {year} {2010})}\BibitemShut {NoStop}%
\bibitem [{\citenamefont {Deng}\ \emph {et~al.}(2013)\citenamefont {Deng},
  \citenamefont {Otto},\ and\ \citenamefont
  {Lupascu}}]{:/content/aip/journal/jap/114/5/10.1063/1.4817512}%
  \BibitemOpen
  \bibfield  {author} {\bibinfo {author} {\bibfnamefont {C.}~\bibnamefont
  {Deng}}, \bibinfo {author} {\bibfnamefont {M.}~\bibnamefont {Otto}}, \ and\
  \bibinfo {author} {\bibfnamefont {A.}~\bibnamefont {Lupascu}},\ }\href
  {\doibase http://dx.doi.org/10.1063/1.4817512} {\bibfield  {journal}
  {\bibinfo  {journal} {Journal of Applied Physics}\ }\textbf {\bibinfo
  {volume} {114}},\ \bibinfo {eid} {054504} (\bibinfo {year}
  {2013})}\BibitemShut {NoStop}%
\bibitem [{\citenamefont {Holder}\ \emph {et~al.}(2013)\citenamefont {Holder},
  \citenamefont {Osborn}, \citenamefont {Lobb},\ and\ \citenamefont
  {Musgrave}}]{PhysRevLett.111.065901}%
  \BibitemOpen
  \bibfield  {author} {\bibinfo {author} {\bibfnamefont {A.~M.}\ \bibnamefont
  {Holder}}, \bibinfo {author} {\bibfnamefont {K.~D.}\ \bibnamefont {Osborn}},
  \bibinfo {author} {\bibfnamefont {C.~J.}\ \bibnamefont {Lobb}}, \ and\
  \bibinfo {author} {\bibfnamefont {C.~B.}\ \bibnamefont {Musgrave}},\ }\href
  {\doibase 10.1103/PhysRevLett.111.065901} {\bibfield  {journal} {\bibinfo
  {journal} {Phys. Rev. Lett.}\ }\textbf {\bibinfo {volume} {111}},\ \bibinfo
  {pages} {065901} (\bibinfo {year} {2013})}\BibitemShut {NoStop}%
\bibitem [{\citenamefont {Gordon}\ \emph {et~al.}(2014)\citenamefont {Gordon},
  \citenamefont {Abu-Farsakh}, \citenamefont {Janotti},\ and\ \citenamefont
  {Van~de Walle}}]{gordon2014hydrogen}%
  \BibitemOpen
  \bibfield  {author} {\bibinfo {author} {\bibfnamefont {L.}~\bibnamefont
  {Gordon}}, \bibinfo {author} {\bibfnamefont {H.}~\bibnamefont {Abu-Farsakh}},
  \bibinfo {author} {\bibfnamefont {A.}~\bibnamefont {Janotti}}, \ and\
  \bibinfo {author} {\bibfnamefont {C.~G.}\ \bibnamefont {Van~de Walle}},\
  }\href@noop {} {\bibfield  {journal} {\bibinfo  {journal} {Scientific
  reports}\ }\textbf {\bibinfo {volume} {4}} (\bibinfo {year}
  {2014})}\BibitemShut {NoStop}%
\bibitem [{\citenamefont {DuBois}\ \emph {et~al.}(2013)\citenamefont {DuBois},
  \citenamefont {Per}, \citenamefont {Russo},\ and\ \citenamefont
  {Cole}}]{PhysRevLett.110.077002}%
  \BibitemOpen
  \bibfield  {author} {\bibinfo {author} {\bibfnamefont {T.~C.}\ \bibnamefont
  {DuBois}}, \bibinfo {author} {\bibfnamefont {M.~C.}\ \bibnamefont {Per}},
  \bibinfo {author} {\bibfnamefont {S.~P.}\ \bibnamefont {Russo}}, \ and\
  \bibinfo {author} {\bibfnamefont {J.~H.}\ \bibnamefont {Cole}},\ }\href
  {\doibase 10.1103/PhysRevLett.110.077002} {\bibfield  {journal} {\bibinfo
  {journal} {Phys. Rev. Lett.}\ }\textbf {\bibinfo {volume} {110}},\ \bibinfo
  {pages} {077002} (\bibinfo {year} {2013})}\BibitemShut {NoStop}%
\bibitem [{\citenamefont {Zeng}\ \emph {et~al.}(2015)\citenamefont {Zeng},
  \citenamefont {Nik}, \citenamefont {Greibe}, \citenamefont {Krantz},
  \citenamefont {Wilson}, \citenamefont {Delsing},\ and\ \citenamefont
  {Olsson}}]{BarrierVary}%
  \BibitemOpen
  \bibfield  {author} {\bibinfo {author} {\bibfnamefont {L.~J.}\ \bibnamefont
  {Zeng}}, \bibinfo {author} {\bibfnamefont {S.}~\bibnamefont {Nik}}, \bibinfo
  {author} {\bibfnamefont {T.}~\bibnamefont {Greibe}}, \bibinfo {author}
  {\bibfnamefont {P.}~\bibnamefont {Krantz}}, \bibinfo {author} {\bibfnamefont
  {C.~M.}\ \bibnamefont {Wilson}}, \bibinfo {author} {\bibfnamefont
  {P.}~\bibnamefont {Delsing}}, \ and\ \bibinfo {author} {\bibfnamefont
  {E.}~\bibnamefont {Olsson}},\ }\href
  {http://stacks.iop.org/0022-3727/48/i=39/a=395308} {\bibfield  {journal}
  {\bibinfo  {journal} {Journal of Physics D: Applied Physics}\ }\textbf
  {\bibinfo {volume} {48}},\ \bibinfo {pages} {395308} (\bibinfo {year}
  {2015})}\BibitemShut {NoStop}%
\bibitem [{\citenamefont {Stoutimore}\ \emph {et~al.}(2012)\citenamefont
  {Stoutimore}, \citenamefont {Khalil}, \citenamefont {Lobb},\ and\
  \citenamefont {Osborn}}]{Stoutimore}%
  \BibitemOpen
  \bibfield  {author} {\bibinfo {author} {\bibfnamefont {M.~J.~A.}\
  \bibnamefont {Stoutimore}}, \bibinfo {author} {\bibfnamefont {M.~S.}\
  \bibnamefont {Khalil}}, \bibinfo {author} {\bibfnamefont {C.~J.}\
  \bibnamefont {Lobb}}, \ and\ \bibinfo {author} {\bibfnamefont {K.~D.}\
  \bibnamefont {Osborn}},\ }\href {\doibase
  http://dx.doi.org/10.1063/1.4744901} {\bibfield  {journal} {\bibinfo
  {journal} {Applied Physics Letters}\ }\textbf {\bibinfo {volume} {101}},\
  \bibinfo {eid} {062602} (\bibinfo {year} {2012})}\BibitemShut {NoStop}%
\bibitem [{\citenamefont {Phillips}(1972)}]{PhillipsJLTP}%
  \BibitemOpen
  \bibfield  {author} {\bibinfo {author} {\bibfnamefont {W.~A.}\ \bibnamefont
  {Phillips}},\ }\href {\doibase 10.1007/BF00660072} {\bibfield  {journal}
  {\bibinfo  {journal} {Journal of Low Temperature Physics}\ }\textbf {\bibinfo
  {volume} {7}},\ \bibinfo {pages} {351} (\bibinfo {year} {1972})}\BibitemShut
  {NoStop}%
\bibitem [{\citenamefont {Anderson}\ \emph {et~al.}(1972)\citenamefont
  {Anderson}, \citenamefont {Halperin},\ and\ \citenamefont
  {Varma}}]{doi:10.1080/14786437208229210}%
  \BibitemOpen
  \bibfield  {author} {\bibinfo {author} {\bibfnamefont {P.~W.}\ \bibnamefont
  {Anderson}}, \bibinfo {author} {\bibfnamefont {B.~I.}\ \bibnamefont
  {Halperin}}, \ and\ \bibinfo {author} {\bibfnamefont {C.~M.}\ \bibnamefont
  {Varma}},\ }\href {\doibase 10.1080/14786437208229210} {\bibfield  {journal}
  {\bibinfo  {journal} {Philosophical Magazine}\ }\textbf {\bibinfo {volume}
  {25}},\ \bibinfo {pages} {1} (\bibinfo {year} {1972})}\BibitemShut {NoStop}%
\bibitem [{\citenamefont {Grabovskij}\ \emph {et~al.}(2012)\citenamefont
  {Grabovskij}, \citenamefont {Peichl}, \citenamefont {Lisenfeld},
  \citenamefont {Weiss},\ and\ \citenamefont {Ustinov}}]{Grabovskij12102012}%
  \BibitemOpen
  \bibfield  {author} {\bibinfo {author} {\bibfnamefont {G.~J.}\ \bibnamefont
  {Grabovskij}}, \bibinfo {author} {\bibfnamefont {T.}~\bibnamefont {Peichl}},
  \bibinfo {author} {\bibfnamefont {J.}~\bibnamefont {Lisenfeld}}, \bibinfo
  {author} {\bibfnamefont {G.}~\bibnamefont {Weiss}}, \ and\ \bibinfo {author}
  {\bibfnamefont {A.~V.}\ \bibnamefont {Ustinov}},\ }\href {\doibase
  10.1126/science.1226487} {\bibfield  {journal} {\bibinfo  {journal}
  {Science}\ }\textbf {\bibinfo {volume} {338}},\ \bibinfo {pages} {232}
  (\bibinfo {year} {2012})}\BibitemShut {NoStop}%
\bibitem [{\citenamefont {Lisenfeld}\ \emph {et~al.}(2015)\citenamefont
  {Lisenfeld}, \citenamefont {Grabovskij}, \citenamefont {M{\"u}ller},
  \citenamefont {Cole}, \citenamefont {Weiss},\ and\ \citenamefont
  {Ustinov}}]{lisenfeld}%
  \BibitemOpen
  \bibfield  {author} {\bibinfo {author} {\bibfnamefont {J.}~\bibnamefont
  {Lisenfeld}}, \bibinfo {author} {\bibfnamefont {G.~J.}\ \bibnamefont
  {Grabovskij}}, \bibinfo {author} {\bibfnamefont {C.}~\bibnamefont
  {M{\"u}ller}}, \bibinfo {author} {\bibfnamefont {J.~H.}\ \bibnamefont {Cole}},
  \bibinfo {author} {\bibfnamefont {G.}~\bibnamefont {Weiss}}, \ and\ \bibinfo
  {author} {\bibfnamefont {A.~V.}\ \bibnamefont {Ustinov}},\ }\href {\doibase
  10.1038/ncomms7182} {\bibfield  {journal} {\bibinfo  {journal} {Nat.
  Commun.}\ }\textbf {\bibinfo {volume} {6}} (\bibinfo {year} {2015}),\
  10.1038/ncomms7182}\BibitemShut {NoStop}%
\bibitem [{\citenamefont {Khalil}\ \emph {et~al.}(2014)\citenamefont {Khalil},
  \citenamefont {Gladchenko}, \citenamefont {Stoutimore}, \citenamefont
  {Wellstood}, \citenamefont {Burin},\ and\ \citenamefont
  {Osborn}}]{khalil2013landau}%
  \BibitemOpen
  \bibfield  {author} {\bibinfo {author} {\bibfnamefont {M.~S.}\ \bibnamefont
  {Khalil}}, \bibinfo {author} {\bibfnamefont {S.}~\bibnamefont {Gladchenko}},
  \bibinfo {author} {\bibfnamefont {M.~J.~A.}\ \bibnamefont {Stoutimore}},
  \bibinfo {author} {\bibfnamefont {F.~C.}\ \bibnamefont {Wellstood}}, \bibinfo
  {author} {\bibfnamefont {A.~L.}\ \bibnamefont {Burin}}, \ and\ \bibinfo
  {author} {\bibfnamefont {K.~D.}\ \bibnamefont {Osborn}},\ }\href {\doibase
  10.1103/PhysRevB.90.100201} {\bibfield  {journal} {\bibinfo  {journal} {Phys.
  Rev. B}\ }\textbf {\bibinfo {volume} {90}},\ \bibinfo {pages} {100201}
  (\bibinfo {year} {2014})}\BibitemShut {NoStop}%
\bibitem [{\citenamefont {Burin}\ \emph {et~al.}(2013)\citenamefont {Burin},
  \citenamefont {Khalil},\ and\ \citenamefont
  {Osborn}}]{PhysRevLett.110.157002}%
  \BibitemOpen
  \bibfield  {author} {\bibinfo {author} {\bibfnamefont {A.~L.}\ \bibnamefont
  {Burin}}, \bibinfo {author} {\bibfnamefont {M.~S.}\ \bibnamefont {Khalil}}, \
  and\ \bibinfo {author} {\bibfnamefont {K.~D.}\ \bibnamefont {Osborn}},\
  }\href {\doibase 10.1103/PhysRevLett.110.157002} {\bibfield  {journal}
  {\bibinfo  {journal} {Phys. Rev. Lett.}\ }\textbf {\bibinfo {volume} {110}},\
  \bibinfo {pages} {157002} (\bibinfo {year} {2013})}\BibitemShut {NoStop}%
\bibitem [{\citenamefont {Sarabi}\ \emph {et~al.}(2015)\citenamefont {Sarabi},
  \citenamefont {Ramanayaka}, \citenamefont {Burin}, \citenamefont
  {Wellstood},\ and\ \citenamefont {Osborn}}]{sarabi2014cavity}%
  \BibitemOpen
  \bibfield  {author} {\bibinfo {author} {\bibfnamefont {B.}~\bibnamefont
  {Sarabi}}, \bibinfo {author} {\bibfnamefont {A.~N.}\ \bibnamefont
  {Ramanayaka}}, \bibinfo {author} {\bibfnamefont {A.~L.}\ \bibnamefont
  {Burin}}, \bibinfo {author} {\bibfnamefont {F.~C.}\ \bibnamefont
  {Wellstood}}, \ and\ \bibinfo {author} {\bibfnamefont {K.~D.}\ \bibnamefont
  {Osborn}},\ }\href {\doibase http://dx.doi.org/10.1063/1.4918775} {\bibfield
  {journal} {\bibinfo  {journal} {Applied Physics Letters}\ }\textbf {\bibinfo
  {volume} {106}},\ \bibinfo {eid} {172601} (\bibinfo {year}
  {2015})}\BibitemShut {NoStop}%
\bibitem [{\citenamefont {Sandberg}\ \emph {et~al.}(2012)\citenamefont
  {Sandberg}, \citenamefont {Vissers}, \citenamefont {Kline}, \citenamefont
  {Weides}, \citenamefont {Gao}, \citenamefont {Wisbey},\ and\ \citenamefont
  {Pappas}}]{Sandberg}%
  \BibitemOpen
  \bibfield  {author} {\bibinfo {author} {\bibfnamefont {M.}~\bibnamefont
  {Sandberg}}, \bibinfo {author} {\bibfnamefont {M.~R.}\ \bibnamefont
  {Vissers}}, \bibinfo {author} {\bibfnamefont {J.~S.}\ \bibnamefont {Kline}},
  \bibinfo {author} {\bibfnamefont {M.}~\bibnamefont {Weides}}, \bibinfo
  {author} {\bibfnamefont {J.}~\bibnamefont {Gao}}, \bibinfo {author}
  {\bibfnamefont {D.~S.}\ \bibnamefont {Wisbey}}, \ and\ \bibinfo {author}
  {\bibfnamefont {D.~P.}\ \bibnamefont {Pappas}},\ }\href@noop {} {\bibfield
  {journal} {\bibinfo  {journal} {Applied Physics Letters}\ }\textbf {\bibinfo
  {volume} {100}},\ \bibinfo {eid} {262605} (\bibinfo {year}
  {2012})}\BibitemShut {NoStop}%
\bibitem [{\citenamefont {Khalil}\ \emph {et~al.}(2013)\citenamefont {Khalil},
  \citenamefont {Stoutimore}, \citenamefont {Gladchenko}, \citenamefont
  {Holder}, \citenamefont {Musgrave}, \citenamefont {Kozen}, \citenamefont
  {Rubloff}, \citenamefont {Liu}, \citenamefont {Gordon}, \citenamefont {Yum},
  \citenamefont {Banerjee}, \citenamefont {Lobb},\ and\ \citenamefont
  {Osborn}}]{ALDOxides}%
  \BibitemOpen
  \bibfield  {author} {\bibinfo {author} {\bibfnamefont {M.~S.}\ \bibnamefont
  {Khalil}}, \bibinfo {author} {\bibfnamefont {M.~J.~A.}\ \bibnamefont
  {Stoutimore}}, \bibinfo {author} {\bibfnamefont {S.}~\bibnamefont
  {Gladchenko}}, \bibinfo {author} {\bibfnamefont {A.~M.}\ \bibnamefont
  {Holder}}, \bibinfo {author} {\bibfnamefont {C.~B.}\ \bibnamefont
  {Musgrave}}, \bibinfo {author} {\bibfnamefont {A.~C.}\ \bibnamefont {Kozen}},
  \bibinfo {author} {\bibfnamefont {G.}~\bibnamefont {Rubloff}}, \bibinfo
  {author} {\bibfnamefont {Y.~Q.}\ \bibnamefont {Liu}}, \bibinfo {author}
  {\bibfnamefont {R.~G.}\ \bibnamefont {Gordon}}, \bibinfo {author}
  {\bibfnamefont {J.~H.}\ \bibnamefont {Yum}}, \bibinfo {author} {\bibfnamefont
  {S.~K.}\ \bibnamefont {Banerjee}}, \bibinfo {author} {\bibfnamefont {C.~J.}\
  \bibnamefont {Lobb}}, \ and\ \bibinfo {author} {\bibfnamefont {K.~D.}\
  \bibnamefont {Osborn}},\ }\href {\doibase
  http://dx.doi.org/10.1063/1.4826253} {\bibfield  {journal} {\bibinfo
  {journal} {Applied Physics Letters}\ }\textbf {\bibinfo {volume} {103}},\
  \bibinfo {eid} {162601} (\bibinfo {year} {2013})}\BibitemShut {NoStop}%
\bibitem [{\citenamefont {Anderson}(1978)}]{anderson1978lectures}%
  \BibitemOpen
  \bibfield  {author} {\bibinfo {author} {\bibfnamefont {P.~W.}\ \bibnamefont
  {Anderson}},\ }\href@noop {} {\bibfield  {journal} {\bibinfo  {journal}
  {Ill-Condensed Matter/La Matiere Mal Condensee, 31 st Session of the Les
  Houches Summer School}\ ,\ \bibinfo {pages} {159}} (\bibinfo {year}
  {1978})}\BibitemShut {NoStop}%
\bibitem [{\citenamefont {Lupa\ifmmode~\mbox{\c{s}}\else \c{s}\fi{}cu}\ \emph
  {et~al.}(2009)\citenamefont {Lupa\ifmmode~\mbox{\c{s}}\else \c{s}\fi{}cu},
  \citenamefont {Bertet}, \citenamefont {Driessen}, \citenamefont {Harmans},\
  and\ \citenamefont {Mooij}}]{PhysRevB.80.172506}%
  \BibitemOpen
  \bibfield  {author} {\bibinfo {author} {\bibfnamefont {A.}~\bibnamefont
  {Lupa\ifmmode~\mbox{\c{s}}\else \c{s}\fi{}cu}}, \bibinfo {author}
  {\bibfnamefont {P.}~\bibnamefont {Bertet}}, \bibinfo {author} {\bibfnamefont
  {E.~F.~C.}\ \bibnamefont {Driessen}}, \bibinfo {author} {\bibfnamefont {C.~J.
  P.~M.}\ \bibnamefont {Harmans}}, \ and\ \bibinfo {author} {\bibfnamefont
  {J.~E.}\ \bibnamefont {Mooij}},\ }\href {\doibase 10.1103/PhysRevB.80.172506}
  {\bibfield  {journal} {\bibinfo  {journal} {Phys. Rev. B}\ }\textbf {\bibinfo
  {volume} {80}},\ \bibinfo {pages} {172506} (\bibinfo {year}
  {2009})}\BibitemShut {NoStop}%
\bibitem [{\citenamefont {Cole}\ \emph {et~al.}(2010)\citenamefont {Cole},
  \citenamefont {M�ller}, \citenamefont {Bushev}, \citenamefont {Grabovskij},
  \citenamefont {Lisenfeld}, \citenamefont {Lukashenko}, \citenamefont
  {Ustinov},\ and\ \citenamefont
  {Shnirman}}]{:/content/aip/journal/apl/97/25/10.1063/1.3529457}%
  \BibitemOpen
  \bibfield  {author} {\bibinfo {author} {\bibfnamefont {J.~H.}\ \bibnamefont
  {Cole}}, \bibinfo {author} {\bibfnamefont {C.}~\bibnamefont {M�ller}},
  \bibinfo {author} {\bibfnamefont {P.}~\bibnamefont {Bushev}}, \bibinfo
  {author} {\bibfnamefont {G.~J.}\ \bibnamefont {Grabovskij}}, \bibinfo
  {author} {\bibfnamefont {J.}~\bibnamefont {Lisenfeld}}, \bibinfo {author}
  {\bibfnamefont {A.}~\bibnamefont {Lukashenko}}, \bibinfo {author}
  {\bibfnamefont {A.~V.}\ \bibnamefont {Ustinov}}, \ and\ \bibinfo {author}
  {\bibfnamefont {A.}~\bibnamefont {Shnirman}},\ }\href {\doibase
  http://dx.doi.org/10.1063/1.3529457} {\bibfield  {journal} {\bibinfo
  {journal} {Applied Physics Letters}\ }\textbf {\bibinfo {volume} {97}},\
  \bibinfo {eid} {252501} (\bibinfo {year} {2010})}\BibitemShut {NoStop}%
\bibitem [{Sup()}]{Supplemental}%
  \BibitemOpen
  \href@noop {} {}\bibinfo {note} {See Supplementary Material [URL will be
  inserted by publisher], which includes Refs.
  \cite{VonSchickfus1977144,gao2008physics}}\BibitemShut {NoStop}%
\bibitem [{\citenamefont {Sarabi}(2014)}]{Bahmanthesis}%
  \BibitemOpen
  \bibfield  {author} {\bibinfo {author} {\bibfnamefont {B.}~\bibnamefont
  {Sarabi}},\ }\emph {\bibinfo {title} {Cavity quantum electrodynamics of
  nanoscale two-level systems}},\ \href@noop {} {Ph.D. thesis},\ \bibinfo
  {school} {University of Maryland - College Park} (\bibinfo {year}
  {2014})\BibitemShut {NoStop}%
\bibitem [{\citenamefont {Stoneking}\ and\ \citenamefont
  {Den~Hartog}(1997)}]{Poisson}%
  \BibitemOpen
  \bibfield  {author} {\bibinfo {author} {\bibfnamefont {M.~R.}\ \bibnamefont
  {Stoneking}}\ and\ \bibinfo {author} {\bibfnamefont {D.~J.}\ \bibnamefont
  {Den~Hartog}},\ }\href@noop {} {\bibfield  {journal} {\bibinfo  {journal}
  {Review of Scientific Instruments}\ }\textbf {\bibinfo {volume} {68}}
  (\bibinfo {year} {1997})}\BibitemShut {NoStop}%
\bibitem [{\citenamefont {Baier}\ and\ \citenamefont
  {Schickfus}(1988)}]{PhysRevB.38.9952}%
  \BibitemOpen
  \bibfield  {author} {\bibinfo {author} {\bibfnamefont {G.}~\bibnamefont
  {Baier}}\ and\ \bibinfo {author} {\bibfnamefont {M.~v.}\ \bibnamefont
  {Schickfus}},\ }\href {\doibase 10.1103/PhysRevB.38.9952} {\bibfield
  {journal} {\bibinfo  {journal} {Phys. Rev. B}\ }\textbf {\bibinfo {volume}
  {38}},\ \bibinfo {pages} {9952} (\bibinfo {year} {1988})}\BibitemShut
  {NoStop}%
\bibitem [{\citenamefont {Kimble}(1998)}]{Kimble}%
  \BibitemOpen
  \bibfield  {author} {\bibinfo {author} {\bibfnamefont {H.~J.}\ \bibnamefont
  {Kimble}},\ }\href {http://stacks.iop.org/1402-4896/1998/i=T76/a=019}
  {\bibfield  {journal} {\bibinfo  {journal} {Physica Scripta}\ }\textbf
  {\bibinfo {volume} {1998}},\ \bibinfo {pages} {127} (\bibinfo {year}
  {1998})}\BibitemShut {NoStop}%
\bibitem [{\citenamefont {Schickfus}\ and\ \citenamefont
  {Hunklinger}(1977)}]{VonSchickfus1977144}%
  \BibitemOpen
  \bibfield  {author} {\bibinfo {author} {\bibfnamefont {M.~V.}\ \bibnamefont
  {Schickfus}}\ and\ \bibinfo {author} {\bibfnamefont {S.}~\bibnamefont
  {Hunklinger}},\ }\href {\doibase
  http://dx.doi.org/10.1016/0375-9601(77)90558-8} {\bibfield  {journal}
  {\bibinfo  {journal} {Physics Letters A}\ }\textbf {\bibinfo {volume} {64}},\
  \bibinfo {pages} {144 } (\bibinfo {year} {1977})}\BibitemShut {NoStop}%
\bibitem [{\citenamefont {Gao}(2008)}]{gao2008physics}%
  \BibitemOpen
  \bibfield  {author} {\bibinfo {author} {\bibfnamefont {J.}~\bibnamefont
  {Gao}},\ }\emph {\bibinfo {title} {The physics of superconducting microwave
  resonators}},\ \href@noop {} {Ph.D. thesis},\ \bibinfo  {school} {California
  Institute of Technology} (\bibinfo {year} {2008})\BibitemShut {NoStop}%
\end{thebibliography}
\end{document}